\documentclass{aa501}
\usepackage{graphicx}

\begin{document}

   \title{A Three Dimensional
   Magnetohydrodynamic Pulse in a Transversely Inhomogeneous Medium}

   \author{D. Tsiklauri and V.M. Nakariakov}

   \offprints{David Tsiklauri, \\ \email{tsikd@astro.warwick.ac.uk}}

   \institute{Physics Department, University of Warwick, Coventry,
   CV4 7AL, U.K. }

   \date{Received ???? 2002 / Accepted ???? 2002}

\abstract{Interaction of impulsively generated MHD waves with a
one-dimensional plasma inhomogeneity, transverse to the magnetic
field, is considered in the three-dimensional regime. 
Because of the transverse inhomogeneity, MHD fluctuations, 
even if they do not include initially any density perturbation, evolve 
toward states where the compressible components tend to become predominant.
The propagating MHD pulse asymptotically
reaches a quasi-steady state with the final levels of density
perturbation weakly depending on the degree of non-planeness of
the pulse in the homogeneous transverse direction and somewhat
stronger depending on plasma $\beta$. Our study demonstrates the
necessity of incorporation of compressible and 3D effects in
theory of Alfv\'en wave phase mixing. However, as far as the
dynamics of weakly non-plane Alfv\'en waves is concerned it can
still be qualitatively understood in terms of the previous 2.5D
models.  }
\titlerunning{A 3D Magnetohydrodynamic Pulse ...}
\authorrunning{Tsiklauri \& Nakariakov}
\maketitle

\section{Introduction}
Problems of heating of the open corona of the Sun and acceleration
of the solar wind are closely related with the interaction of MHD
waves with plasma inhomogeneities and, in particular, with
coupling of compressible and incompressible components of the
waves. Investigation of coupling mechanisms between the  waves is
of prime importance because, on one hand, the Alfv\'en waves are
usual candidates of energy transport from the lower layers of the
solar atmosphere to the corona, and, on the other, compressible
perturbations are subject to much more efficient dissipation than
the incompressible Alfv\'en waves (see, e.g., \cite{nu96}). Also,
the compressive waves, as opposed to the Alfv\'en waves, can
transport energy across the magnetic field, due to the fact that
their propagation in space is not constrained by the field lines.

The original idea of the phase mixing of {\it incompressible}
Alfv\'en waves (\cite{hp83}) was based on the following argument:
when plasma has a density gradient perpendicular to the magnetic
field, local Alfv\'en speed is a function of the transverse
coordinate. Thus, when an Alfv\'en wave propagates along the field
its perturbations on the adjacent field lines become out of phase.
This stretching of Alfv\'en wave front creates progressively
smaller spatial scales across the field. In turn, because the
dissipation is proportional to the wave number squared, phase
mixing leads to enhanced dissipation of the Alfv\'en wave
directly. In the {\it compressible} plasma, as demonstrated by
\cite{malara}; \cite{nrm97}; \cite{nrm98}; \cite{Botha};
\cite{td1}; \cite{td2}, phase mixing of linearly polarized plane
Alfv\'en waves leads to the enhanced nonlinear generation of fast
magnetoacoustic waves. However, it was established, in 2.5D
geometry, for the harmonic Alfv\'en wave \cite{Botha}, and for a
wide spectrum Alfv\'en pulse by \cite{td1} that compressive
perturbations, which are initially absent from the system, do not
grow to a substantial fraction of the initial Alfv\'en wave due
the destructive wave interference effect.

In this  work we consider  fully three dimensional geometry and
study interaction of linear (with the non-linear effects totally
ignored) MHD waves with a one-dimensional inhomogeneity of the
plasma, taking into account {\it compressibility} of the plasma
and the {\it localization} of the MHD pulse in the direction
perpendicular to both the magnetic field and the inhomogeneity
gradient. In particular, we shall study how the phenomenon of
phase mixing is affected by these factors. A similar problem was
studied by \cite{gg00} in the context of resonant absorption of
MHD waves in coronal loops as a heating mechanism. However, our
treatment is more general as we perform a direct numerical 3D
simulation without resorting to Fourier transform in time and $y$
direction (and, consequently, our study can easily be generalized
to the case of 2D and 3D structuring) allowing us to consider an
initial value problem. Besides, \cite{gg00} used a harmonic form
of the initial perturbation as opposed to our spatially localized,
Gaussian, one. Recently, \cite{hbw02} showed that the wide
spectrum regime of phase mixing can be quite different from the
harmonic one, as in fact, phase mixing of localized Alfv\'en
pulses results in a slower, algebraic, damping as opposed to the
standard exponential damping of harmonic Alfv\'en waves, which
suggest that localized Alfv\'enic perturbations will transport
energy higher into the corona than harmonic ones. Such Alfv\'enic
perturbations could be generated e.g. by transient events such as
solar flares, coronal mass ejections, etc. While most of the
phase-mixing studies concentrated on harmonic perturbations (e.g.
\cite{hp83}; \cite{hip97}; \cite{hgi97}; \cite{nrm98};
\cite{rnr98}; \cite{dhia99}; \cite{dha00}; \cite{glb00};
\cite{Botha}), only few considered spatially localized ones
(\cite{nrm97}; \cite{td1}; \cite{td2}; \cite{hbw02}).

An additional motivation to this study is connected with the
growing interest to the problem of interaction of MHD waves with
2D and 3D plasma structures and irregularities, as the coronal and
wind plasmas are observed to be structured in all three
dimensions. 2D and 3D structuring can dramatically affect
properties of MHD waves. In particular, as it was shown in the
incompressible regime by \cite{ss89} and confirmed in numerical
experiments performed by \cite{pmv98} and \cite{mpv00}, the
structuring dramatically increases the efficiency of wave
dissipation. However, in those studies,  the compressible effects 
were not taken into account. 
From this point of view, our
investigation of the interaction of a 3D MHD pulse with a
1D plasma inhomogeneity in the compressible regime provides a
necessary building element of the general theory of MHD wave
interaction with plasma inhomogeneities.

The present model is based upon the MHD description of plasma and we
focus on the problems that are of relevance to the heating of
solar corona, acceleration and the dynamics of solar wind. While,
similar studies exist that deal with, for instance, terrestrial
auroral applications. Namely, an alternative to MHD, more relevant,
particle-in-cell simulations 
(using a guiding center implicit code for the electrons)
 have been performed (\cite{glq99}; \cite{gml00};
\cite{glm01}), which study the propagation and
collisionless dissipation (via effective electron beam generation)
of the Alfv\'en waves in the auroral density inhomogeneities
(cavities).

In this work we study the propagation part of the problem, i.e. we consider
an ideal plasma limit. The work is in progress to include finite 
plasma resistivity in order to investigate quantitatively the
dependence of the decay of Alfv\'enic part of the MHD pulse 
upon the coupling to the existing compressive  waves.

The paper is organized as follows: in section 2 we
formulate our model. In section 3 we present the results of
numerical simulation, while we close in section 4 with the
discussion of main results.

\section{The model}

\begin{figure*}
\centering
 \includegraphics[width=17cm]{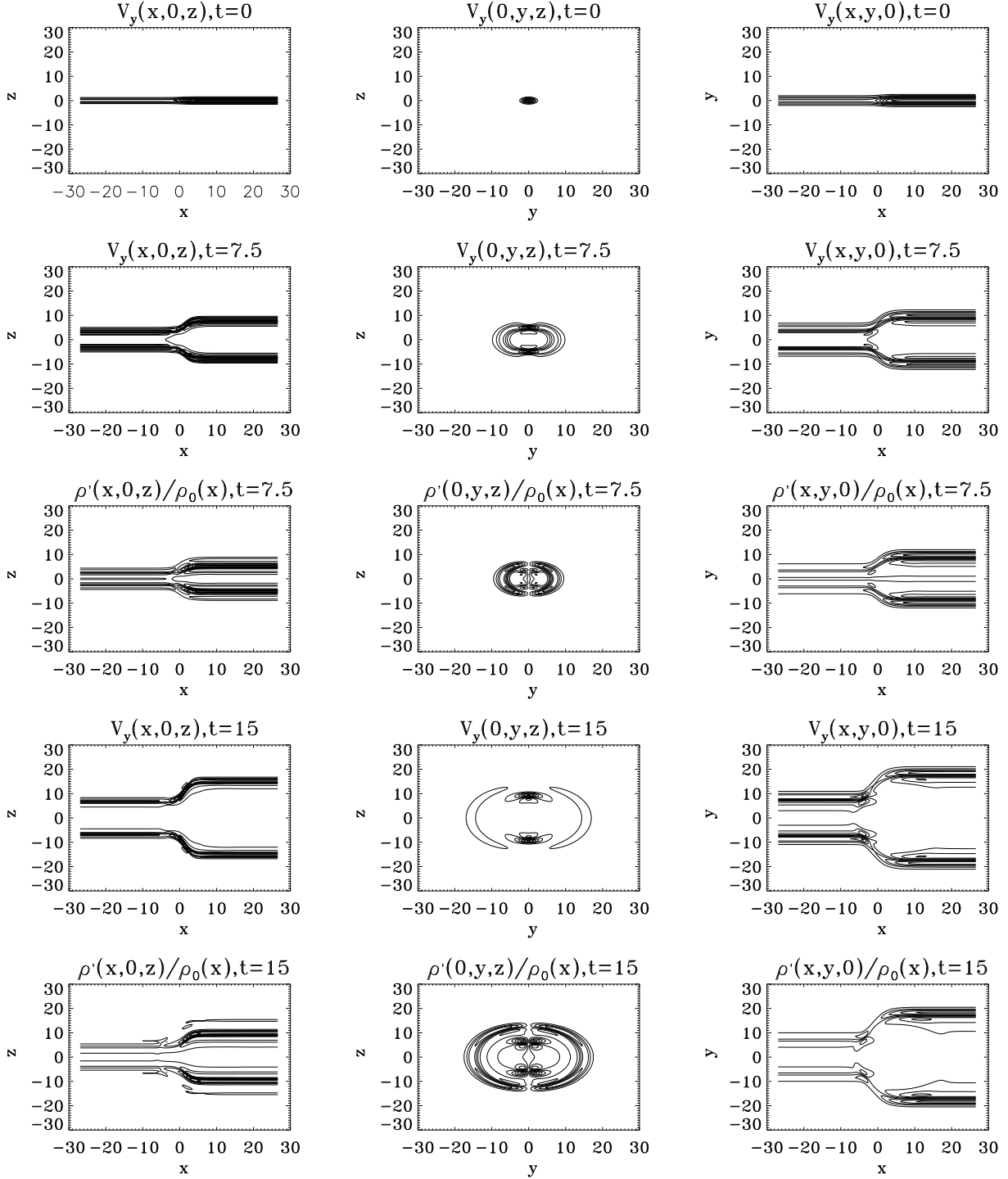}
\caption{Contour plots of $V_y$ and
$\rho'/\rho_0(x)$ (relative density perturbation)
through three cross-sections at
the same time snapshots. Here, $\beta=0.5$, $\alpha_y=0.6$, $\alpha_z=1.0$.}
\label{vy}
\end{figure*}

In our model we use equations  ideal MHD
\begin{equation}
\rho {{\partial \vec V}\over{\partial t}} +
\rho(\vec V \cdot \nabla) \vec V = - \nabla p -{{1}\over{4 \pi}}
\vec B \times {\rm curl} \vec B, \label{1}
\end{equation}
\begin{equation}
{{\partial \vec B}\over{\partial t}}= {\rm curl} (\vec V \times \vec B),
\label{2}
\end{equation}
\begin{equation}
{{\partial p}\over{\partial t}} + \vec V \cdot \nabla p + \gamma p
\nabla \cdot \vec V=0,
\label{3}
\end{equation}
where $\vec B$ is the magnetic field, $\vec V$ is plasma velocity,
$\rho$ is plasma mass density, and $p$ is plasma thermal pressure.
In what follows we use $5/3$ for the value of $\gamma$.

We solve equations (\ref{1})-(\ref{3}) in Cartesian coordinates ($x,y,z$).
Note that as we solve a fully 3D problem we retain
 variation in the
$y$-direction, i.e. ($\partial / \partial y \not = 0$). The equilibrium
state is taken to be an inhomogeneous plasma of density $\rho_0(x)$
and a uniform magnetic field $B_0$ in the $z$-direction.
We consider a plasma configuration similar to the
one investigated in \cite{malara}; \cite{nrm97}; \cite{nrm98};
\cite{Botha}; \cite{td1}, i.e. the plasma
has a one-dimensional inhomogeneity in the equilibrium density
$\rho_0(x)$ and temperature $T_0(x)$.
The unperturbed thermal pressure, $p_0$, is taken to be constant
everywhere.

Next, we do usual linearization of the Eqs.~(\ref{1})-(\ref{3})
and write them in component form as following
\begin{equation}
\rho_0(x){{\partial V_x}\over{\partial t}} +
{{\partial p}\over{\partial x}} -
{{B_0}\over{4 \pi}}
\left( {{\partial  B_x}\over{\partial z}}-
{{\partial  B_z}\over{\partial x}} \right) =0,
\label{5}
\end{equation}
\begin{equation}
\rho_0(x){{\partial V_y}\over{\partial t}} +
{{\partial p}\over{\partial y}} +
{{B_0}\over{4 \pi}}
\left( {{\partial  B_z}\over{\partial y}}-
{{\partial  B_y}\over{\partial z}} \right) =0,
\label{6}
\end{equation}
\begin{equation}
\rho_0(x){{\partial V_z}\over{\partial t}} +
{{\partial p}\over{\partial z}} =0,
\label{7}
\end{equation}
\begin{equation}
{{\partial B_x}\over{\partial t}}-
B_0{{\partial V_x}\over{\partial z}}=0,
\label{8}
\end{equation}
\begin{equation}
{{\partial B_y}\over{\partial t}}-
B_0{{\partial V_y}\over{\partial z}}=0,
\label{9}
\end{equation}
\begin{equation}
{{\partial B_z}\over{\partial t}}+B_0
\left( {{\partial V_x}\over{\partial x}} +
{{\partial V_y}\over{\partial y}}\right)=0,
\label{10}
\end{equation}
\begin{equation}
{{\partial p}\over{\partial t}}+\gamma p_0
\left( {{\partial V_x}\over{\partial x}} +
{{\partial V_y}\over{\partial y}} +
{{\partial V_z}\over{\partial z}}\right)=0.
\label{11}
\end{equation}
It is useful to re-write Eqs.~(\ref{5})-(\ref{11}) in a form
of three coupled wave equations as following
\begin{equation}
\left[ \partial^2_{tt}-\left(c_s^2(x)+c_A^2(x)\right) \partial^2_{xx}
-c_A^2(x) \partial^2_{zz}\right]V_x
\label{12}
\end{equation}
$$
-\left[\left(c_s^2(x)+c_A^2(x)\right) \partial^2_{xy}
\right]V_y-
\left[ c_s^2(x)\partial^2_{xz}
\right]V_z=0,
$$
\begin{equation}
\left[ \partial^2_{tt}-\left(c_s^2(x)+c_A^2(x)\right) \partial^2_{yy}
-c_A^2(x) \partial^2_{zz}\right]V_y
\label{13}
\end{equation}
$$
-\left[\left(c_s^2(x)+c_A^2(x)\right) \partial^2_{xy}
\right]V_x-
\left[ c_s^2(x)\partial^2_{yz}
\right]V_z=0,
$$
\begin{equation}
\left[ \partial^2_{tt}
-c_s^2(x) \partial^2_{zz}\right]V_z
-\left[c_s^2(x) \partial^2_{xz}
\right]V_x
\label{14}
\end{equation}
$$
-
\left[ c_s^2(x)\partial^2_{yz}
\right]V_y=0,
$$
where $c_A(x)=B_0 / \sqrt{4 \pi \rho_0(x)}$ and
$c_s(x)=\sqrt{\gamma p_0 /\rho_0(x)}$ denote local Alfv\'en
and sound speeds respectively.

We solve Eqs.(\ref{5})-(\ref{11}) numerically after re-writing
them in a dimensionless form using following normalization:
$B_{x,y,z}=B_0 {\bar B}_{x,y,z}$, $(x,y,z)=a_* ({\bar x},{\bar
y},{\bar z})$, $c_A(x)=B_0 / \sqrt{4 \pi \rho_0(x)}= B_0 / \sqrt{4
\pi \rho_*}/\sqrt{3-2 \, \tanh(\lambda x)}= c^*_A/ \sqrt{3-2 \,
\tanh(\lambda x)}$, $t=(a_*/c^*_A){\bar t}$, $V_{x,y,z}=c^*_A
{\bar V}_{x,y,z}$. Note, that $c_s(x)= \sqrt{\gamma \beta /2}
c_A(x)$, where $\beta$ stands for the ratio of thermal and
magnetic pressures $\beta=p_0/(B_0^2/8 \pi)$. Here, $\lambda$ is a
free parameter which controls the steepness  of the density
profile gradient. In our simulations we use $\lambda=0.5$. In what
follows we omit bars on top of the physical quantities.

\section{Numerical Results}

In order to solve Eqs.(\ref{5})-(\ref{11}) numerically
we have written a new numerical code {\it dt4dx10},
which uses a high order finite difference scheme.
Namely, it evaluates 10-th order centered spatial
derivatives, and advances solution in time using
4-th order Runge-Kutta algorithm.
Therefore, {\it dt4dx10} is $O[(\Delta t / T)^4]$ accurate
in time and $O[(max(\Delta x,  \Delta y, \Delta z)/ L)^{10}]$
accurate in space, with $T$ and $L$ denoting run
time and linear size of the simulation domain.
The use of such a high-order numerical approach was motivated
by the very nature of the problem considered, as development of phase
mixing leads to the generation of very steep profiles in the wave front.

The simulation cube size is set by the limits $-25.0 \leq x \leq 25.0$,
$-25.0 \leq y \leq 25.0$
and  $-25.0 \leq z \leq 25.0$.
Boundary conditions used in all our simulations are
zero-gradient in all three spatial dimensions.

\begin{figure}[]
\resizebox{\hsize}{!}{\includegraphics{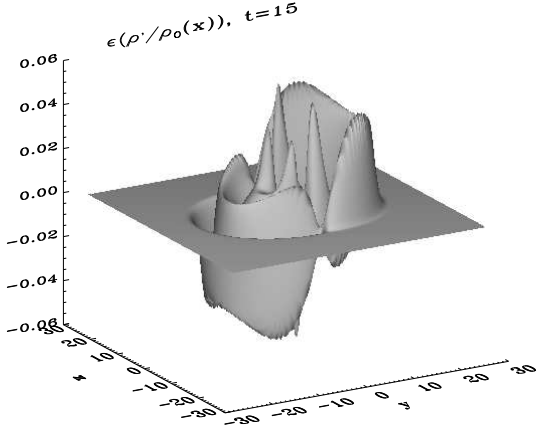}} \caption{A
snapshot of relative perturbation of plasma density by an MHD
pulse interacting with a one-dimensional inhomogeneity at $t=15$
through $x=0$ cross-section. Here, $\beta=0.5$, $\alpha_y=0.6$,
$\alpha_z=1.0$.} \label{rhox}
\end{figure}

\begin{figure}[]
\resizebox{\hsize}{!}{\includegraphics{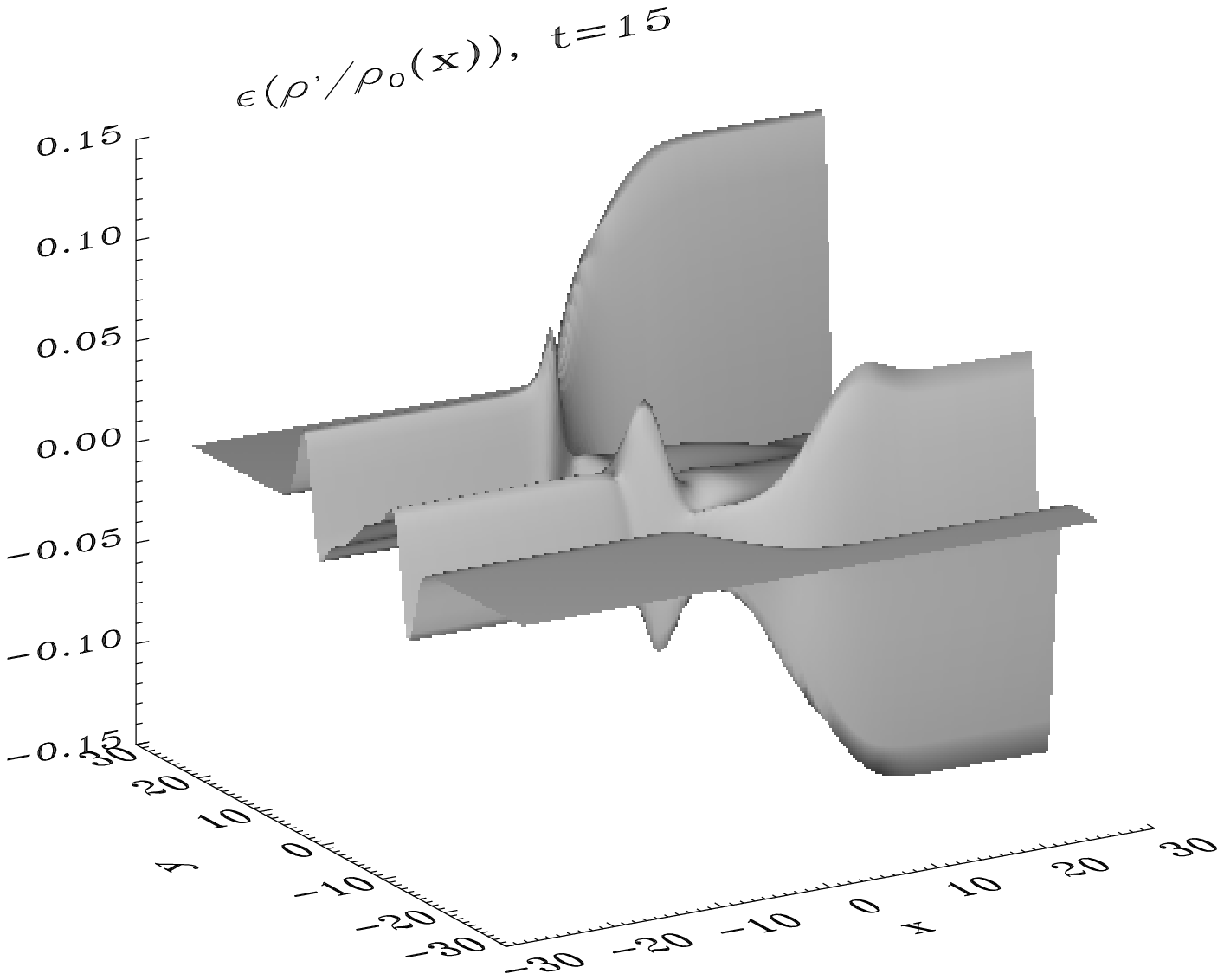}} \caption{The
same as in Fig.~(\ref{rhox}), but through $z=0$ cross-section.}
\label{rhoz}
\end{figure}

\begin{figure}[]
\resizebox{\hsize}{!}{\includegraphics{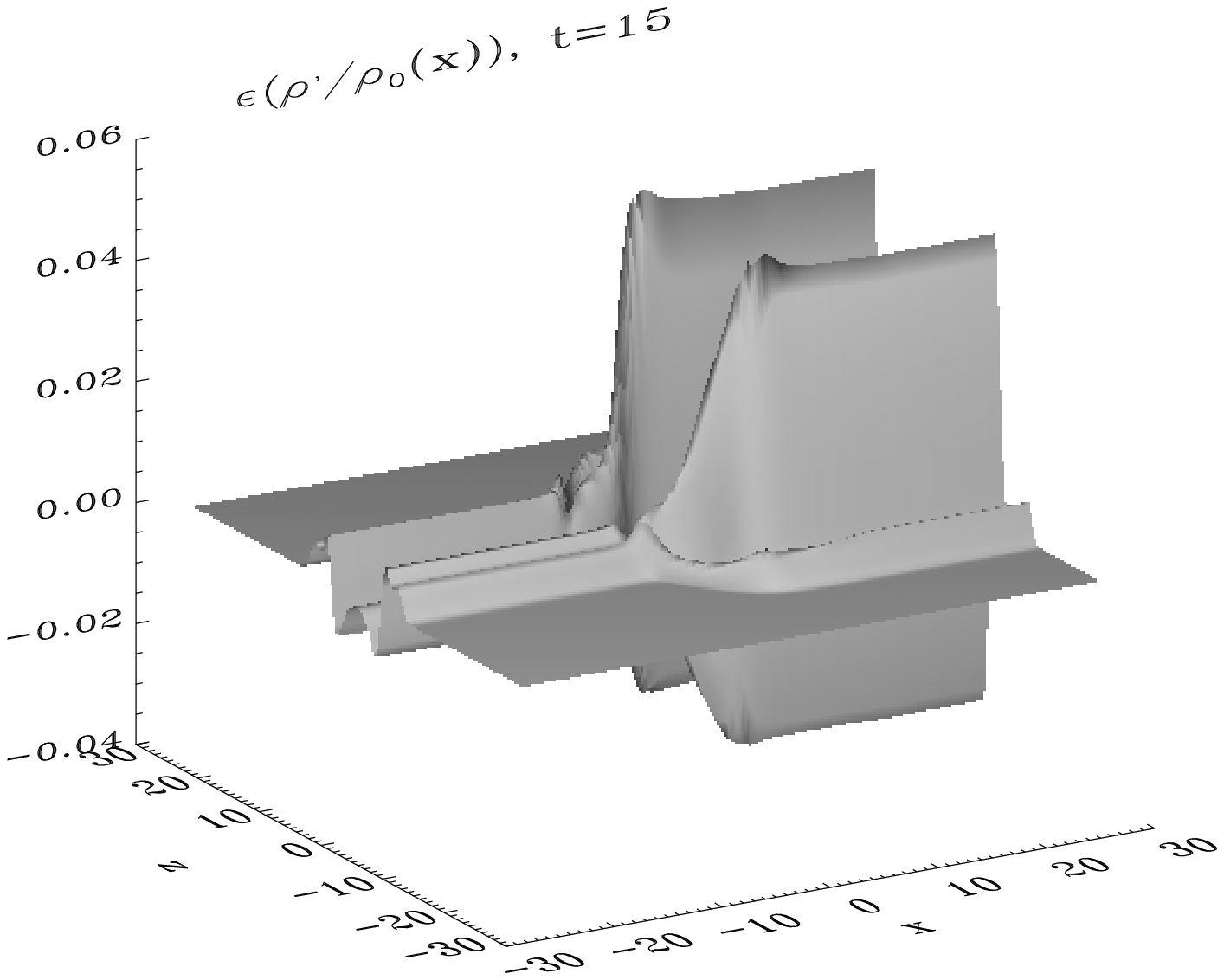}} \caption{The
same as in Fig.~(\ref{rhox}), but through $y=0$ cross-section.}
\label{rhoy}
\end{figure}

\begin{figure*}
\centering
 \includegraphics[width=17cm]{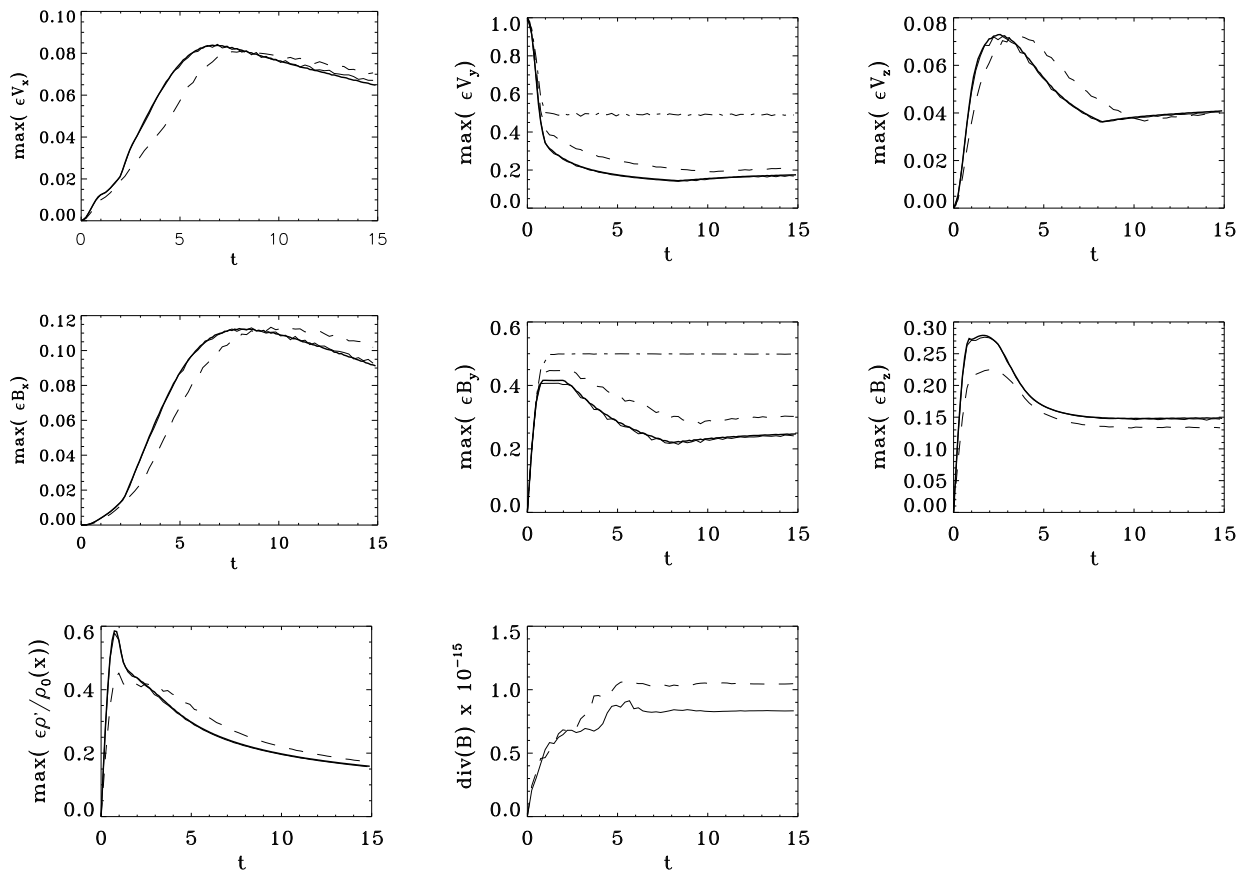}
\caption{Maximum of the absolute values over the whole 3D
simulation box of all physical quantities as a function of time.
Here, $\beta=0.5$ and $\alpha_z=1.0$. Thick solid lines correspond
to the resolution $256^3$, while thin solid lines to that of
$128^3$, both for $\alpha_y=0.6$. Note that on all seven panels
these two curves practically overlap, which serves as a proof of
convergence of our simulation results. Dashed lines represent
$\alpha_y=0.4$, while dash-dotted lines show $\alpha_y=0$.}
\label{max_all}
\end{figure*}

We have performed calculation on various resolutions in
attempt to achieve convergence of the results.
The graphical results presented here are for the spatial
resolution $128^3$, which refers to number of
grid points in $x,y$ and $z$ directions respectively.
We have also performed calculation on the spatial resolution
$256^3$ and we found that the
results converge perfectly. This is understandable
due to the high order of the scheme even for the
$128^3$ resolution
time-error, $O[(\Delta t / T)^4]\approx 7 \times 10^{-8}$, and
spatial-error, $O[(max(\Delta x,  \Delta y, \Delta z)/ L)^{10}]
\approx 8 \times 10^{-22}$.
Since a $256^3$ resolution run takes about 8 hours
on Compaq ES40 with eight EV6 500-MHz processors,
while a $128^3$ resolution
takes only 1 h on Compaq ES40 with four EV6 500-MHz processors
producing the same results, we opted for the
latter, less CPU-consuming, alternative.

In the numerical simulations the MHD perturbation is initially a
plane (with respect to $x$-coordinate) pulse, which has a Gaussian
structure in  $y$ and $z$-coordinates
\begin{equation}
V_y(x,y,z,t=0)=c_A(x) \exp \left(-(\alpha_y y)^2-(\alpha_z z)^2\right).
\label{15}
\end{equation}
Here, $\alpha_y$ and $\alpha_z$ are free parameters which control
the strength of gradients in $y$ and $z$ direction of the initial
perturbation. As the problem considered is linear, the wave
amplitude can be taken to be normalized to unity.

In the geometry considered, when the initial perturbation depends
on $y$ it cannot be regarded as pure Alfv\'enic one. In fact, in
such a pulse, all three waves -- Alfv\'en, fast and slow
magnetosonic  waves -- are inter-coupled so that there is no use of
their separation {\it per se}. The pulse is set to be initially
plane in the $x$-direction. This allows us to emphasize the effect
of the inhomogeneity on the pulse evolution. From the point of
view of applications, this simply means that the initial
characteristic size of the pulse in that direction is greater than
the scale of the inhomogeneity.

 The particular choice of the initial
condition Eq.~(\ref{15}) is motivated by the argument that when we
set $B_y$ initially to zero we automatically guarantee fulfillment
of $\mathrm{div}\vec B=0$ when $\alpha_y \not = 0$, and as the
system evolves it adjusts itself which mode to excite (depending
whether $\alpha_y$ is zero or not). For instance, if we choose our
initial conditions as $V_y$ given by Eq.~(\ref{15}),
$B_y(x,y,z,t=0)= \exp \left(-(\alpha_y y)^2-(\alpha_z
z)^2\right)$, with $\alpha_y=0$ and the rest of physical
quantities set to zero, we would excite a pure Alfv\'enic pulse
traveling in the negative direction along $z$-axis. However, as
long as $\alpha_y \not = 0$ we have to set $B_y(x,y,z,t=0)=0$ in
order to fulfill $\mathrm{div} \vec B=0$. Besides, our choice of
the localized \lq\lq kinematic" (the velocity is initially
perturbed only, while the magnetic field is constant) perturbation
Eq.~(\ref{15}) is well motivated by the fact that such
perturbations can arise, e.g., during coronal mass ejections,
solar flares, or other violent events (see, e.g., Roussev et al.
2001), etc.

\begin{figure}[]
\resizebox{\hsize}{!}{\includegraphics{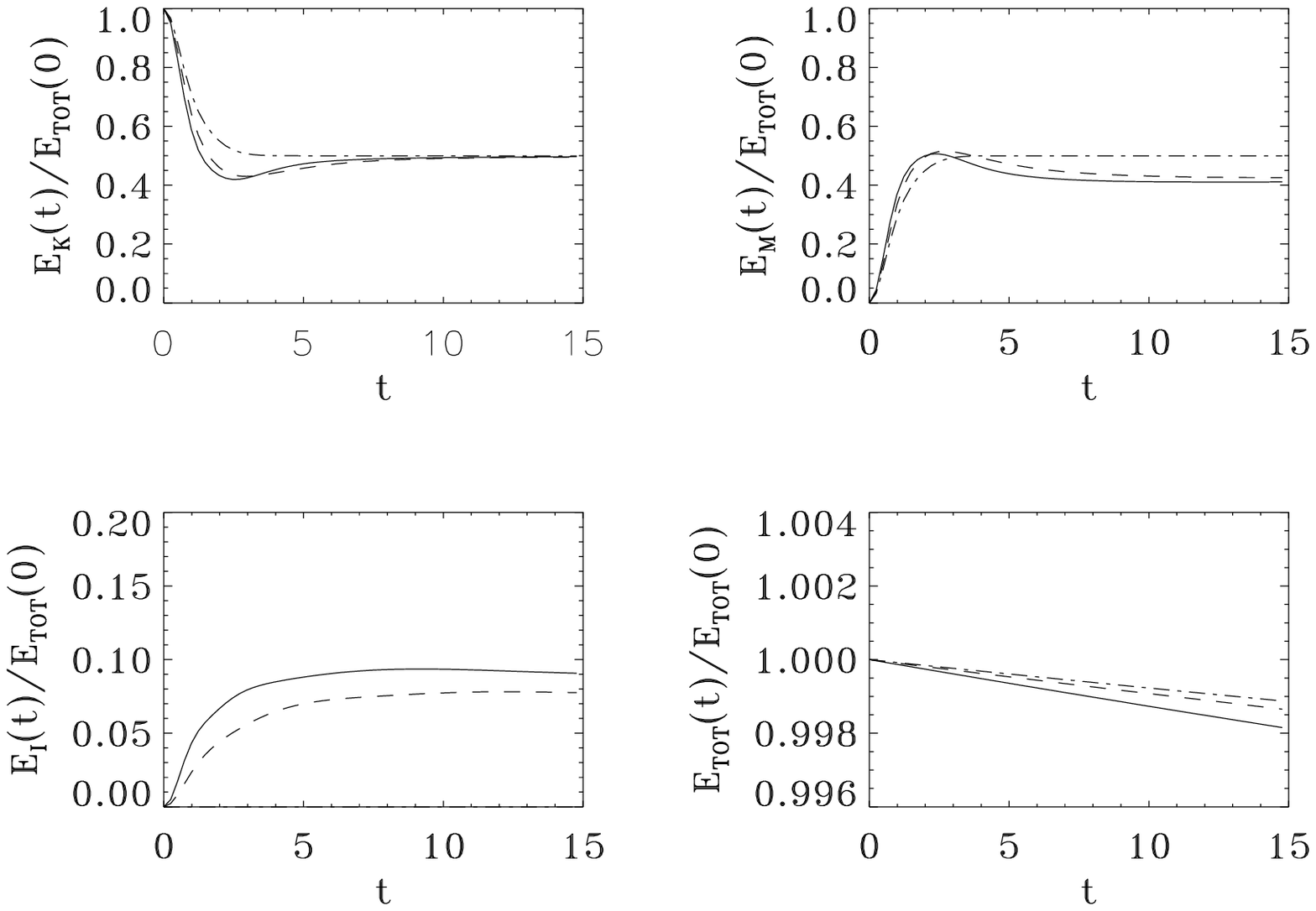}}
\caption{Normalized kinetic, magnetic, internal and total energies
(all of them integrated over the volume)
as a function of time. Here, $\beta=0.5$ and $\alpha_z=1.0$.
Solid lines correspond to $\alpha_y=0.6$, while dashed and
dash-dotted lines show cases of $\alpha_y=0.4$ and $\alpha_y=0$
respectively.}
\label{eng}
\end{figure}

\begin{figure*}
\centering
 \includegraphics[width=17cm]{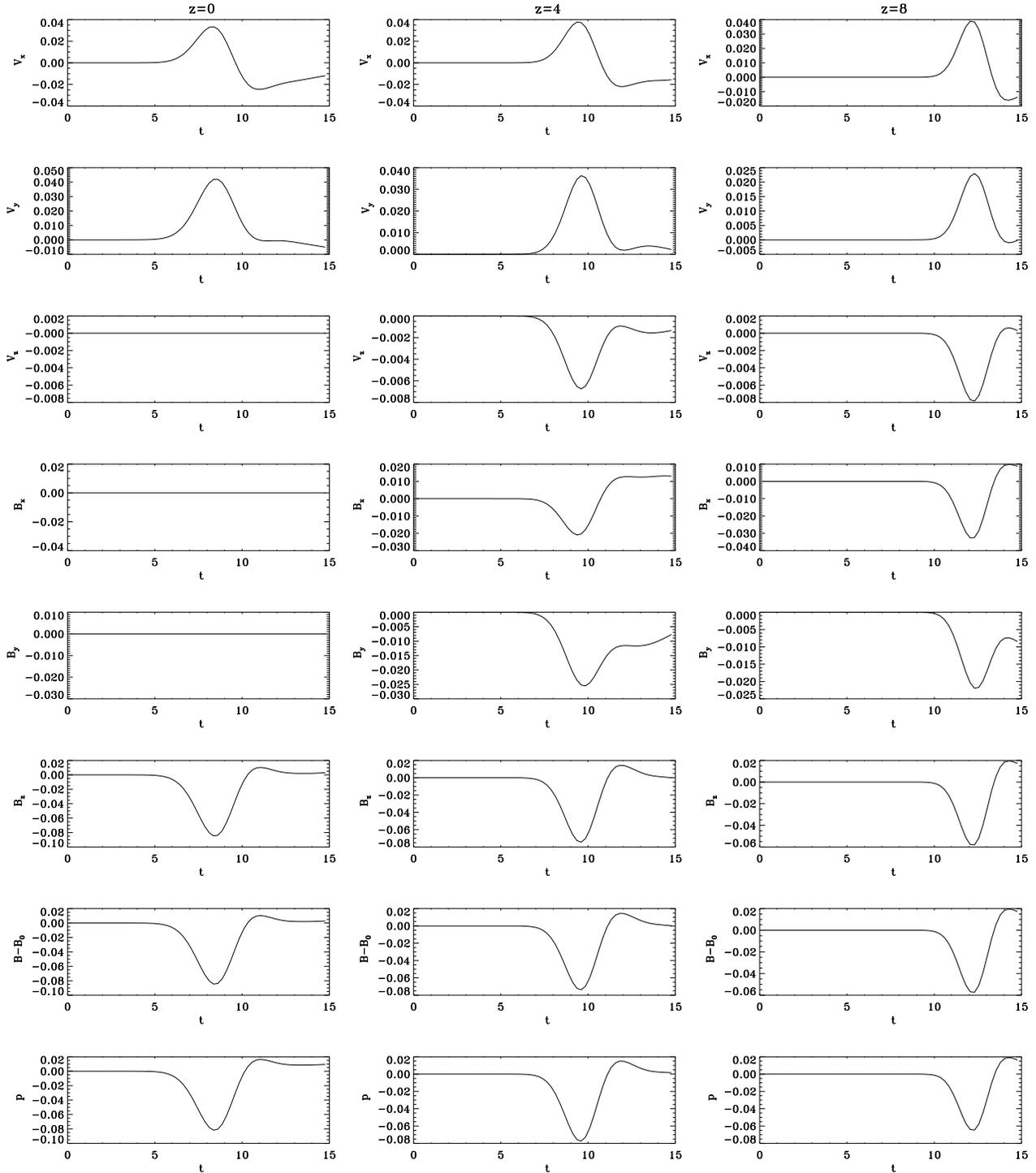}
\caption{All physical quantities perturbed by the MHD pulse as a
function of time at a given point in the phase mixing region.
Here, $\beta=0.5$, $\alpha_y=0.6$, $\alpha_z=1.0$. Left column: at
the point (x=0,y=-8,z=0), mid column: at the point (x=0,y=-8,z=4),
right column at the point (x=0,y=-8,z=8)} \label{x65y45}
\end{figure*}

\begin{figure*}
\centering
 \includegraphics[width=17cm]{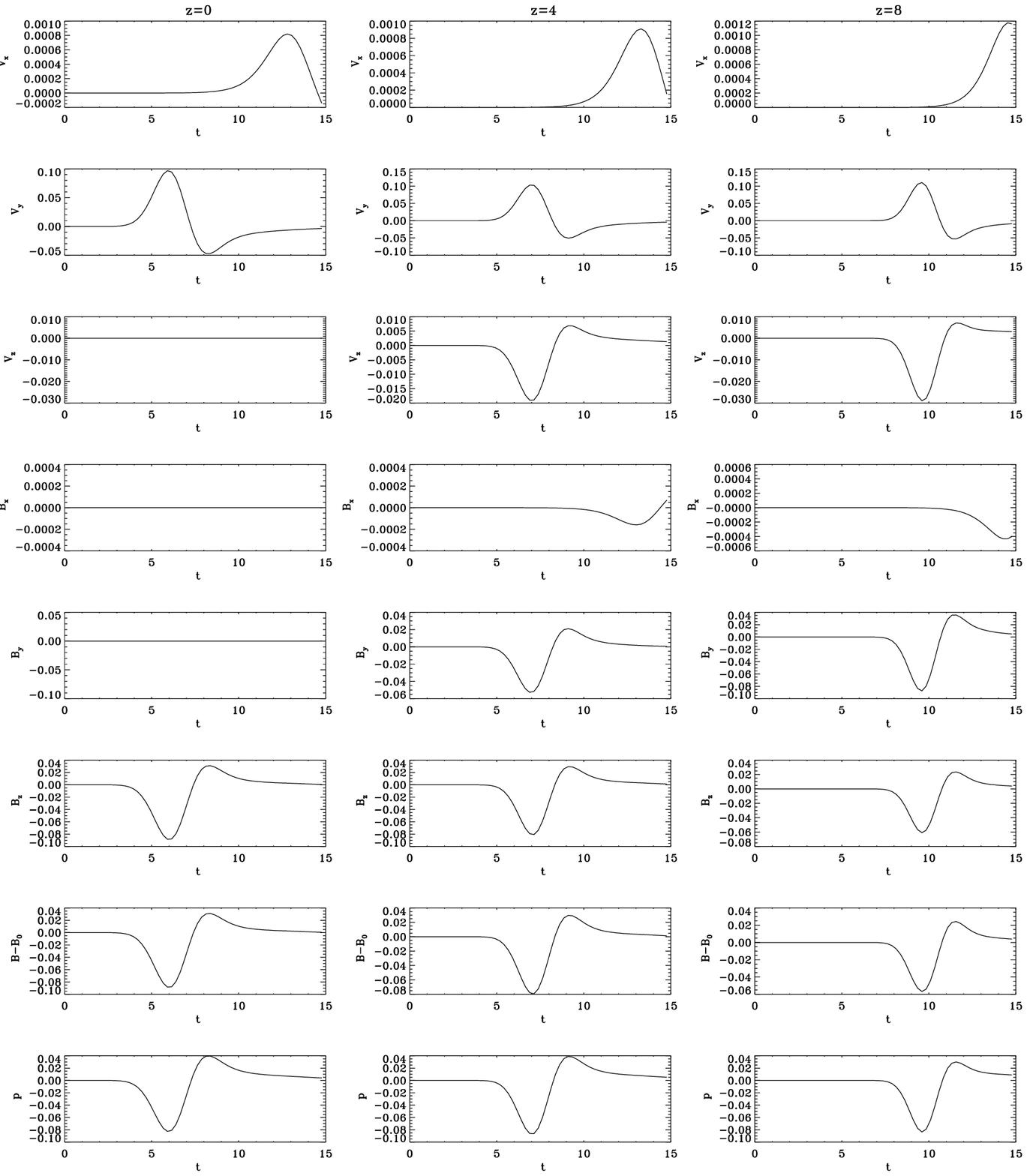}
\caption{All physical quantities as a function of time at a given
point far from the phase mixing region. Here, $\beta=0.5$,
$\alpha_y=0.6$, $\alpha_z=1.0$. Left column: at the point
(x=16,y=-8,z=0), mid column: at the point (x=16,y=-8,z=4), right
column at the point (x=16,y=-8,z=8)} \label{x105y45}
\end{figure*}

In Figure~\ref{vy}  we present time evolution of the transverse
component of the plasma velocity, $V_y$ and the density
perturbation $\rho'/\rho_0(x)$. The choice of these physical
quantities is motivated by the following reasoning: the transverse
velocity component $V_y$ is subject to Alfv\'en wave phase mixing
and the density perturbation demonstrates the effect of the plasma
compressibility. The left column of Fig.~\ref{vy} shows three
snapshots of $V_y$ and two snapshots ($\rho'/\rho_0(x)$) through
the $y=0$ cross-section, the middle column shows three snapshots
of $V_y$ and two snapshots ($\rho'/\rho_0(x)$) through the $x=0$
cross-section, and the right column shows three snapshots of $V_y$
and two snapshots ($\rho'/\rho_0(x)$) through the $z=0$
cross-section. The snapshots of $V_y$ and  $\rho'/\rho_0(x)$ at
times $t=7.5$ and 15 are placed such that it is easier to spot the
differences. The snapshot of $\rho'/\rho_0(x)$ at $t=0$ is not
present in the figure as it is identically zero and it is
generated later by the $y$-gradients in the function $V_y$. This
compressible perturbation is a significant component of the pulse.
What we gather from this graph is as following:
\begin{itemize}
\item $y=0$ cross-section: the initial MHD pulse,
which is plane with respect to $x$-coordinate, is split in two
D'Alambert's solutions, with half of the amplitudes, traveling in
two opposite directions along the magnetic field. Because the
unperturbed density is inhomogeneous across the magnetic field (in
the $x$-direction) the local Alfv\'en speed depends upon $x$.
Thus, the Alfv\'enic pulse is phase-mixed and its initially plane
front is continuously distorted creating transverse gradients.
Note, that the right wing of the pulse is asymmetric to the left
one, this is due to the fact that its amplitude drops from 1.0 in
the $x>0$ domain to $1/\sqrt{5}$ in the $x<0$. The generated
density perturbation evolves in the similar manner, however there
is a notable difference -- as plasma $\beta=0.5$ in this case, we
observe that the density perturbation travels at a slower speed
(slow magnetosonic wave speed) than the Alfv\'enic one. If we look
at the right wing of $V_y(x,0,z)$ at $t=15$, we see that it
traveled to z=15, which is consistent with the Alfv\'en speed
equals 1 (recall that for $x>0$ $c_A(x)=1$). The density
perturbation as it travels along the field should have a velocity
of a slow magnetosonic wave, which for the field aligned
propagation coincides with the local speed of sound. For the
parameters considered $c_s(x)= \sqrt{\gamma \beta /2} c_A(x)=0.65
c_A(x)$. Therefore, the right wing of the density perturbation at
$t=15$ should have traveled to a position $z \approx 9.7$. That
is what we actually see from the bottom panel of the left column
in Fig.{\ref{vy}}.
\item $x=0$ cross-section: the initial Alfv\'enic pulse
which has an asymmetry as $\alpha_y \not = \alpha_z$ Gaussian
bell shape is split in two smaller amplitude bells traveling in
two opposite directions along the magnetic field plus some
semi-elliptic shaped wake expanding outwards in all directions.
Note, that because  the unperturbed density is homogeneous across
the magnetic field in $y$-direction we do not see any distortions
on the wave front with respect to the $y=0$ line as we saw in the
previous case. The generated density perturbation evolves in the
similar manner, however a notable difference (can be better seen
from Fig.~\ref{rhox}) is that the solution consists of positive
and negative spikes plus semi-elliptic shaped wakes that are
symmetric with respect to $\rho'/\rho_0(x)=0$ plane. This can be
understood readily as due to the conservation of total mass,
integral of density perturbation over the volume should, indeed,
be zero.
\item $z=0$ cross-section: the MHD pulse evolution,
viewed through this cross-section, is similar to the case of $y=0$
cross-section (see above). Again we see two phase-mixed
D'Alambert's solutions propagating in the opposite directions. The
notable difference is that through $z=0$ cross-section the pulse
looks wider as $\alpha_y=0.6$ while $\alpha_z=1.0$. The generated
density perturbation evolves in the similar manner, as the
Alfv\'enic one. However, in differ from $y=0$ cross-section, we
see no difference in traveling speeds. This can be explained as
following: the pulse was not excited as a \lq\lq pure Alfv\'en
wave". In fact, should the perturbation $V_y$ be purely Alfv\'enic
there would have been no propagation along $y$ axis at all (recall
that the magnetic field is directed along the $z$-axis). Thus,
what we see through the $z=0$ cross-section both for $V_y$ and
$\rho'/\rho_0(x)$ is the compressive part of the perturbation
traveling at the fast magnetosonic wave speed. For the parameters
considered, the fast magnetosonic wave speed is $\sqrt{1+\gamma
\beta /2} c_A(x)=1.2 c_A(x)$. Therefore, the right wing of both
$V_y$ and
 density perturbation at $t=15$
should have traveled to a position $y \approx 17.9$. That is what
we actually observe from the two bottom panels of the right column in
Fig.{\ref{vy}}.
\end{itemize}
Also, the figure shows the obvious anisotropy in the pulse
propagation, connected with the presence of two chosen directions:
the direction of the straight magnetic field and the direction of
the density gradient.

Perturbation of $B_y$ is initially absent from the system, but as
it is a potential component of the kinetic counterpart, $V_y$, of
the wave, it is soon generated and further evolves similarly to
$V_y$ in a form of two negative and positive phase-mixed pulses
traveling into two opposite directions. Also, the development of
the pulse is accompanied by generation of compressible components.
The structure of the density perturbations in the pulse at the
time $t=15$ is shown in Figures.~\ref{rhox}, \ref{rhoz} and
\ref{rhoy}. Note, that in all the figures, pre-factor $\epsilon$
stands to underscore the fact that the linear problem, which we
study, does not depend on the initial amplitude. Thus, we use
$A=1$, while we have to bear in mind that linear approximation
itself is valid for small amplitudes that is why we use small,
arbitrary pre-factor $\epsilon$ in our notations.

A fairly good quantity describing the relation between different
components of the pulse is the maximum of absolute value over the
whole simulation domain. We plot this quantity for all physical
variables in Figure~\ref{max_all}. In this figure, thick solid
lines correspond to the resolution $256^3$, while thin solid lines
to that of $128^3$, both for $\alpha_y=0.6$. It is remarkable that
on all panels these two curves practically do overlap, which
serves as a proof of convergence of our simulation results. The
last figure in the bottom row presents the maximum of absolute
value over the whole simulation domain of $\mathrm{div}\, \vec B$,
and we indeed observe almost perfect fulfillment of the
fundamental law, $\mathrm{div}\,\vec B=0$, which comes as a bonus
of high-order, centered, finite difference numerical scheme. As
expected, Fig.~\ref{max_all} shows that when $\alpha_y=0$, the
pulse is perfectly Alfv\'enic (perturbing $V_y$ and $B_y$ only).
This conclusion can be deducted either from analyzing
Eqs.~(\ref{12})-(\ref{14}) or resorting to a classic mechanical
analogy of coupled pendulums. In effect,
Eqs.~(\ref{12})-(\ref{14}) also describe three inter-coupled
mathematical pendulums which are located in $xOz$ plane. Thus, as
long as we do not perturb these pendulums such that $\alpha_y
\not=0$, they will always oscillate in the $xOz$ plane. Thus, what
we see in Fig.~\ref{max_all} when $\alpha_y=0$ (dash-dotted lines)
is that initial kinematic Alfv\'en perturbation ($V_y$) is split
in half-amplitude D'Alambert's solutions (both for $V_y$ and
$B_y$) and no other physical quantity is generated. However, when
we switch on the coupling, $\alpha_y \not=0$, compressible
perturbations are generated.

In order to investigate the effect of phase mixing on the created
quasi-steady MHD state, we produced time series of all physical
quantities in several points of the simulation  3D box,
Figures~\ref{x65y45} and \ref{x105y45}. Namely, at the points
(x=0,y=-8,z=0, 4, 8), which are located in the phase mixing
(spatial inhomogeneity) domain, and at (x=16,y=-8,z=0, 4, 8),
which are far away from it. The three panels (from the left to the
right), in effect, trace the dynamics in the $z$-direction, i.e.
along the regular magnetic field lines. These types of data would
be obtained by, for example, three satellites which are located in
three different regions of solar wind.

There are two noteworthy features that can be gathered from
these plots: First, we see that in the phase mixing region, $V_x$,
which is associated with the fast magnetosonic component of the
pulse, attains about 40 times larger values than in the region
that is far from the phase mixing region. That is a sensible
result, since it is known that phase mixing efficiently generates
oblique fast magnetosonic waves. Second, the total magnetic field
perturbation $|\vec B- \vec B_0|$ is {\it positively } correlated
with the pressure (as well as density, which is proportional to
the gas pressure) perturbation, which indeed is what was expected
from the fast magnetosonic wave.

Also, Figs~\ref{max_all}-\ref{x105y45} demonstrate that,
interacting with the plasma inhomogeneity, the longitudinally
(parallel to the magnetic field) propagating part of the initial
perturbation
develops to an almost non-evolving (without change in
amplitude) propagating state. In this
quasi-steady MHD state all components of the pulse tend to
propagate without change in amplitude, even though the effect of
phase mixing is in action all the time ($x$-component of the
inverse characteristic spatial scale tends to infinity).

In Figure~\ref{eng} we investigate the energetics of our numerical
simulation. In fact, we observe nearly perfect ($\pm 0.2$\% error)
conservation of total energy by {\it dt4dx10} numerical code. The
major conclusion which can be drawn from this graph is that when
$\alpha_y=0$ there is no internal (compressive) energy generation,
while with the increase of $\alpha_y$ its final, asymptotic levels
increase progressively.

\begin{figure}[]
\resizebox{\hsize}{!}{\includegraphics{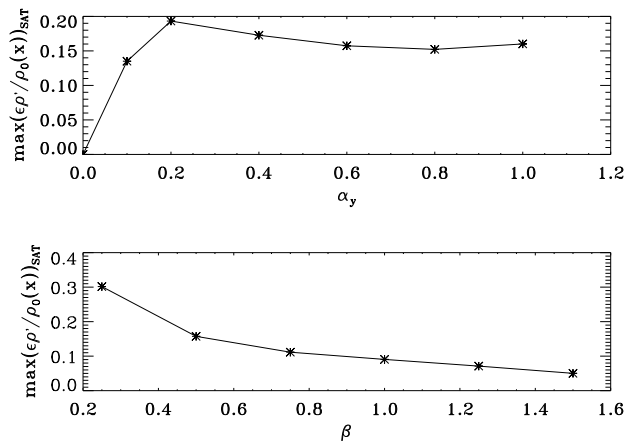}} 
\caption{Top panel:  the final levels of relative density 
perturbations  as a function of $\alpha_y$ 
(for $\beta=0.5$ and $\alpha_z=1.0$). 
Bottom panel: the same
but as a function of $\beta$ 
(for $\alpha_y=0.6$ and $\alpha_z=1.0$).} \label{par}
\end{figure}

The parametric space of the problem is studied in Figure~\ref{par}
by plotting the final levels of relative density perturbation,
approached by the MHD pulse, as a function of the initial
localization of the pulse in the $y$-direction $\alpha_y$, and
plasma $\beta$. We gather from Fig.~\ref{par} that the achieved
levels of density fluctuations depend weakly on $\alpha_y$, while
there is somewhat stronger dependence on $\beta$.

\section{Conclusions}

In this paper we present results of numerical modeling of
interaction of an initially localized MHD pulse with a transverse
inhomogeneity of the plasma.  The pulse is non-plane in all three
spatial directions and is subject to the effect of phase mixing.
An impulsively generated perturbation of the plasma velocity
develops to an anisotropically propagating pulse, which has a
significant compressible component. More specifically:

\begin{itemize}
\item The non-uniformness of the pulse in the homogeneous transverse
(or, in other words, in the third, perpendicular to both the
magnetic field and the inhomogeneity gradient) direction (non-zero
$\alpha_y$) leads to appearance of compressive perturbations in
the pulse.

\item In the presence of plasma inhomogeneity, a non-uniform MHD pulse
is essentially compressible. More specifically, when the pulse is
initially plane in the homogeneous transverse direction,
$\alpha_y=0$, there is no internal (compressive) energy
generation, while with the increase of $\alpha_y$, the levels of
the compressible energy increase progressively.

\item A propagating MHD pulse asymptotically reaches a quasi-steady
state. The final levels of density perturbation, which can be
considered as a measure of the compressibility in the pulse,
depend weakly on $\alpha_y$, while there is somewhat stronger
dependence on $\beta$.

\item
A smooth 1D transverse inhomogeneity of the plasma supports
propagation of {\it compressible} MHD pulses. This mechanism of
wave guiding is different from the well-understood phenomenon of
refraction of fast magnetosonic waves, as the profile of the
Alfv\'en speed in the inhomogeneity does not have a minimum.

\item
Weakly non-plane MHD waves are subject to phase mixing and can
qualitatively be considered in terms of the 2.5D theory of
Alfv\'en wave phase mixing. Consequently, the 2.5D theory of
Alfv\'en wave phase mixing remains relevant to the 3D case too.
However, quantitative theories of the interaction of MHD waves
with plasma inhomogeneities should include the compressibility of
the plasma as a necessary ingredient.

\end{itemize}

Our main conclusion is that the effects of {\it three-dimensionality,
compressibility and inhomogeneity} should be all together 
taken into account in
the wave-based theories of coronal heating and solar wind
acceleration, and as well as in the theories of MHD turbulence.

\begin{acknowledgements}
The authors are grateful to Tony Arber for valuable advise
when {\it dt4dx10} code was written and for a number of valuable discussions.
DT acknowledges financial support from PPARC.
Numerical calculations of this work were
performed using the PPARC funded Compaq MHD Cluster at St Andrews
and Astro-Sun cluster at Warwick.
\end{acknowledgements}

\end{document}